# Surface-Plasmon-Polariton Laser based on an Open-Cavity Fabry-Perot Resonator

Wenqi Zhu[1,2], Ting Xu[1,2†], Amit Agrawal[1,2] and Henri J. Lezec[1*]

[1]Center for Nanoscale Science and Technology, National Institute of Standards and Technology, Gaithersburg, MD 20899, USA.

[2]Maryland Nanocenter, University of Maryland, College Park, MD, 20742 USA.

[†]Current address: National Laboratory of Solid State Microstructures, College of Engineering and Applied Sciences and Collaborative Innovation Center of Advanced Microstructures, Nanjing University, 22 Hankou Road, Nanjing 210093, China.

*Corresponding author. E-mail: henri.lezec@nist.gov

**Abstract: Recent years have witnessed growing interest in the development of small-footprint lasers for potential applications in small-volume sensing and on-chip optical communications. Surface-plasmons – electromagnetic modes evanescently confined to metal-dielectric interfaces – offer an effective route to achieving lasing at nanometer-scale dimensions when resonantly amplified in contact with a gain-medium. Here, we achieve visible frequency ultra-narrow linewidth lasing at room-temperature by leveraging surface plasmons propagating in an open Fabry-Perot cavity formed by a flat metal surface coated with a subwavelength-thick layer of optically-pumped gain medium and orthogonally bound by a pair of flat metal sidewalls. Low perturbation transmission-configuration sampling of the lasing plasmon mode is achieved via an evanescently coupled recessed nanoslit, opening the way to high-figure-of-merit refractive-index sensing of analytes interacting with the open cavity.**

Ongoing miniaturization of the footprint of optical systems and circuits requires development of wavelength scale optical elements able to efficiently generate, manipulate and detect light. Achieving coherent optical sources with narrow linewidth and nanoscale mode-confinement in one or more dimensions has been central to this quest(*1*). Nanolasers incorporating high-index dielectric resonators under the form of photonic crystals (*2–4*) or whispering-gallery-mode cavities (*5–7*) have constituted a major focus of research over the last few decades. More recently, metal based resonators sustaining surface plasmons have emerged as a promising alternative to achieving lasing at nanometer-scale dimensions (*8*, *9*). For example, optically-pumped lasing plasmon modes formed by localized surface plasmons on dye-coated resonant nanoparticles and nanoparticle arrays (*10–12*), or by gap plasmons confined to a dielectric layer of nanoscale thickness between a metal surface and a semiconductor gain medium (*13–17*), have been recently demonstrated. Here, we show, for the first time, lasing of propagating surface plasmon polaritons (SPPs) confined to an open Fabry-Perot cavity. Ultra-narrow linewidth, visible frequency lasing is achieved at room-temperature. The cavity consists of a template-stripped ultra-smooth flat Ag surface orthogonally bound by a pair of Ag sidewalls, forming a trough-shaped, micron-scale width resonator for SPPs exhibiting a record-high quality factor. In particular, the cavity is designed to operate at the red end of the visible spectrum, targeting a nominal resonant operation wavelength $\lambda_0$ = 650 nm. By incorporating a subwavelength thickness dye-doped polymer film on the trough floor, and optically pumping the dye from the open side of the cavity, we demonstrate room-temperature SPP lasing in the red with a linewidth ($\approx$ 0.25 nm) that is substantially narrower than that of localized–plasmon or gap–plasmon lasers demonstrated to date. Transmission-configuration sampling of the lasing intensity with minimum perturbation to the SPP cavity mode is achieved via an evanescently coupled nanoslit recessed below the cavity floor, opening the way to high-figure-of-merit refractive-index sensing of analytes interacting with the open cavity.

A flat surface of noble metal facing a dielectric medium forms a canonical medium for sustaining surface plasmon polaritons – transverse-magnetic (TM) polarized electromagnetic waves which propagate along the interface and decay evanescently normal to the interface, both into the metal (typically on a deep-subwavelength scale) and into the dielectric (typically on the scale of a wavelength). The rectangular open-cavity structure conceived here to form the optical resonator of a visible-frequency SPP laser is defined by a planar, optically opaque Ag 'floor' – delimited by two optically opaque, quasi-planar, quasi-parallel Ag 'sidewalls', oriented normal to the floor (Figure 1A). White-light illumination of the cavity launches SPPs along the metal-air interface of the floor via scattering at the sidewall-floor corners. These SPPs in-turn propagate across the cavity and reflect efficiently from the sidewalls which act like SPP mirrors (*18*, *19*). The cavity is able to sustain resonance of SPP modes propagating normal to the mirrors (in the *x*-direction), for which mirror separation *l* corresponds to an integer number of SPP half wavelengths. The rectangular profile of the resonant cavity allows for SPP standing waves of arbitrary extent parallel to the mirrors (along the *y*-direction), depending only on the specific excitation geometry. In order to avoid scattering losses of SPP field intensities at the top corner of the trough-shaped cavity, the sidewall mirror height *h* is chosen to exceed SPP decay length *d* normal to the cavity floor (in the *z*-direction). A slit fabricated in the cavity floor, of subwavelength width (in the *x*-direction) and arbitrarily length (in the *y*-direction), evanescently connects the cavity to a transparent substrate situated on its reverse side, which allows probing of the cavity in a convenient free-space transmission configuration. Orienting the slit parallel to the cavity sidewalls – in other words parallel to the magnetic field vector of the SPP standing wave – allows efficient coupling between the SPP field in the cavity and the TM-polarized fundamental propagating mode of the slit.

The frequency dependent quality-factor of the cavity resonator is directly related to the round-trip loss of the trapped SPP mode, which includes (i) the decay length of the SPPs along the cavity floor due to absorption and scattering from the metal film, (ii) the SPP reflection losses due scattering and absorption at each of the metal sidewalls, (iii) the length of the Fabry-Perot cavity $l$, and (iv) the scattering loss from the recessed sampling slit.

First (i), to minimize absorption-induced SPP propagation loss in the red, the cavity is fabricated using Ag for the metal surfaces of the cavity. The propagation loss is further minimized by use of ultra-smooth template-stripped Ag (*20*), yielding here surfaces of root-mean-square roughness ≈ 2.9 nm (Supplementary Figure S1). The corresponding SPP propagation decay length is experimentally characterized to be ≈ 25 μm at $\lambda_0 = 650$ nm (Supplementary Figure S2), which is an order of magnitude larger than that expected for a typical as-deposited Ag surface.

Second (ii), performing finite-difference time-domain (FDTD) simulations over the visible frequency range, the reflectance $R_\infty$ of an SPP mode impinging on an infinitely tall, perfectly flat Ag mirror is analyzed as a function of angular deviation $\Delta\theta$ of the mirror with respect to the normal to the propagation-plane (Figure 1B). The SPP reflectance at $\lambda_0 = 650$ nm is maximum at normal impingence ($\Delta\theta = 0°$), yielding $R_\infty = 0.95$, and remains virtually constant (within the range $0.93 \leq R_\infty \leq 0.95$) for $|\Delta\theta| \leq 1°$, an angular deviation tolerance found to be feasible with fabrication sequence used in this study (Supplementary Section I). The natural evanescent confinement of SPPs normal to the direction of propagation furthermore suggests the possibility of reducing the sidewall mirror heights to finite values such that scattering loss at the top of the sidewall remains minimal. FDTD simulations at $\lambda_0 = 650$ nm, assuming perfectly vertical sidewalls, are used to study the fundamental dependence of sidewall reflectivity $R$ on sidewall height $h$. Simulations show that $R$ remains within 1 % of its asymptotic limit for an infinitely tall sidewall for sidewall heights $h > 1.5$ µm (Supplementary Figure S3), in other words for heights exceeding by a factor of $\approx 4$ the corresponding SPP decay length given by simulation ($d \approx 400$ nm).

Third (iii), given a specific round-trip SPP loss, the length of the SPP Fabry Perot cavity governs the tradeoff between the highest achievable quality factor $Q$ and Finesse $\mathcal{F}$ of a given resonance mode (Supplementary Section V). Here, we choose a cavity length $l = 2.5$ µm to yield a resonance peak near a target free-space wavelength of $\lambda_0 = 650$ nm. An FDTD simulation of a cavity having this length, vertical sidewall mirrors of height $h = 2$ µm, and metal surfaces of permittivity corresponding to template-stripped Ag, yields a resonance peak at free-space wavelength $\lambda_{peak} = 647$ nm with simultaneously high values of quality factor and Finesse, respectively $Q_0 = 375$ and $\mathcal{F} = 24$ (Supplementary Figure S4). The resonance condition for the chosen value of $l$ corresponds to the trapping of eight SPP half wavelengths, based on an SPP wavelength $\lambda_{SPP} = 625$ nm derived from the experimentally measured complex permittivity value of template stripped Ag at $\lambda_0 = 647$ nm ($\varepsilon_{Ag} = -16 + 0.6i$).

Finally (iv), the slit is engineered to couple to the field of the SPP standing wave with a minimum level of perturbation to the cavity resonance while still providing acceptable levels of transmission for observation. To achieve this, the slit is fabricated such that it does not intersect the plane of the cavity floor, but instead is recessed below the resonator floor by a distance $\delta$ (Figure 1C), establishing an intentionally weak evanescent link with the SPP mode trapped in the cavity. Moreover, the width of the slit at its closest point approaching to the cavity floor, $\sigma$, is simultaneously optimized for minimal perturbation. The dependence of $Q$ on slit recess $\delta$ and end-width $\sigma$ is explored using FDTD simulations, conservatively assuming a slit taper angle of 15° and a slit position laterally offset to the left of the center of the cavity by a distance $\zeta = 155$ nm. The simulations show that $Q$ monotonically increases as a function of increasing recess $\delta$ and decreasing end-width $\sigma$. Consequently, the slit end-width is nominally set to $\sigma = 10$ nm, the smallest value anticipated to be achievable within the fabrication constraints of the present study. Simultaneously choosing a recess $\delta = 50$ nm yields a quality factor $Q = 350$, which is close to the maximum achievable value for a slit-free (unperturbed) cavity, $Q_0 = 375$. The simulated quality factor $Q$ (Figure 1D) exhibits a periodic dependence on the lateral slit-position $\zeta$, varying 180° out-of-phase with the surface magnetic field intensity, $|H_y|^2$, sampled at an equivalent position $\zeta$ along the floor of a slit-less, otherwise identical reference cavity illuminated under identical conditions. $Q$ is observed to take on local maximum values at lateral slit-positions corresponding to the magnetic field nodes of the reference SPP standing wave, confirming minimum recessed-slit-induced scattering loss at such locations. The chosen slit offset of $\zeta = 155$ nm satisfies this condition.

The FDTD simulated magnetic field $H_y$ of the optimized slit-decorated cavity at resonance ($\lambda_0 = 647$ nm, Figure 1E) is characterized by a laterally trapped SPP standing wave in the cavity, evanescently decaying normal to the cavity floor, and for which one of the magnetic field nodes coincides with the location of the recessed slit. The slit samples a small-fraction of the time-varying electromagnetic field at the surface of the cavity floor by coupling to the $x$-directed electric field component of the SPP standing wave, also minimal at this position, converting it into a guided-mode, and radiating it into the far-field on the back-side of the cavity. The spectrum of the transmitted intensity, monitored at a distance 500 nm below the exit-aperture of the slit, reveals a cavity mode with a resonance located at $\lambda_{peak}$ = 647 nm characterized by a full-width half-maximum (FWHM) linewidth Δ = 1.8 nm, corresponding to a high quality factor $Q = \lambda_{peak} / \Delta$ = 360 (Figure 1F).

Cavity resonator fabrication (Supplementary Section I) is performed using the template stripping method (20). The pattern template, fabricated using electron-beam lithography and cryogenic deep-Si etching, consists of a reusable, high-aspect-ratio Si mesa (Figure 2A) having a flat ultra-smooth top surface (corresponding to the original polished surface of the constituent Si wafer) along with sidewalls that are orthogonal to within an angular tolerance of ≤ 0.2°, over a vertical distance of ≈ 3 µm from the top surface. As the first step in the replication process, Ag is angle-evaporated onto the rotating Si template, to top-surface and sidewall thicknesses of ≈ 300 nm and ≈ 250 nm, respectively. A slit with a nominal lateral slit offset of 155 nm relative to the center of the mesa is then patterned through the top-surface Ag-film via low-current focused-ion-beam (FIB) milling, stopping short of the Ag-Si interface by a recess distance $\delta$ ≈ 50 nm, with an ultra-narrow slit end-width $\sigma$ ≈ 10 nm and a natural slit sidewall taper angle of ≈ 5°. Template stripping then yields an open cavity (Figure 2B) having features that conspire to minimize SPP propagation and scattering losses, namely a recessed sampling slit located at $\zeta$ ≈ 155 nm, an ultra-smooth Ag surface for the cavity floor as well as quasi-vertical Ag sidewalls each having an outward-taper angle ≤ 0.2° over a sidewall height $h$ ≈ 3 µm.

The optical properties of a fabricated SPP resonator is characterized by illuminating the open-side of the cavity with incoherent white-light, collecting the light transmitted through the slit using an inverted optical microscope, and dispersing the collected light onto a grating spectrometer (Supplementary Section II). The resulting spectrum of the collected intensity $I_t$ reveals a pronounced resonance peak centered at $\lambda_{peak}$ = 642 nm (close to simulated peak position of 647 nm) and characterized by a linewidth $\Delta = 2.1 \pm 0.1$ nm (Figure 2C). The uncertainty in the measurement of the resonance linewidth is one standard deviation of the Lorentzian fit parameter. The resulting $Q = 310 \pm 15$ is, to our knowledge, the largest quality factor achieved to date using a surface-plasmon resonator of any type. The uncertainty in the calculated $Q$ is one standard deviation based on propagation of uncertainty from the resonance linewidth curve fitting.

Based on the optimized design of a rectangular cavity SPP resonator, we implement a laser based on stimulated amplified emission of SPPs by incorporating a thin layer of solid-state gain medium into the open cavity, and shifting the recessed slit position to a new position $\zeta' = 55$ nm (Figure 3A). The gain medium consists of a host medium of polymethyl methacralate (PMMA) doped with 4-(Dicyanomethylene)-2-methyl-6-(4-dimethylaminostyryl)-4$H$-pyran (DCM) laser dye to a concentration of 3 mM, spin-coated to a thickness of ≈ 260 nm on the cavity floor (Figure 3B, inset). The PMMA:DCM gain medium absorbs photons with a maximum absorption efficiency at $\lambda_P = 480$ nm, and emits light in the wavelength range 560 nm $\lesssim \lambda_E \lesssim$ 715 nm, with a maximum at 627 nm (*21*).

The resulting gain-medium-coated SPP resonator is optically characterized at room temperature by pumping its open-cavity side with a normally-incident tunable nanosecond pulsed laser ($\lambda_P = 480$ nm, repetition rate = 10 Hz, pulse width ≈ 5 ns), and spectrally characterizing the light emitted under the form of cavity SPPs, as sampled by the slit. The presence of distinct peaks in the spectrum of the as-collected emission intensity, $I_E$, as well as their evolution as a function of pump energy density $I_P$, hint at the possibility of SPP lasing (Figure 3B). This hypothesis is substantiated by analyzing the evolution, as a function of increasing $I_P$, of the emission linewidth (Figure 3C) and emission intensity (Figure 3C, inset), respectively, of the pronounced spectral peak observed at $\lambda_E = 637$ nm. Specifically, the evolution of the emission intensity, displaying a canonical 'kink' shape on a log–log scale (Figure 3D), is consistent with three distinct emission regimes characteristic of a standard laser as it transitions through threshold: (i) spontaneous emission for $I_P \lesssim 0.1$ mJ/cm$^2$, (ii) amplified spontaneous emission for $0.1$ mJ/cm$^2 \lesssim I_P \lesssim 0.3$ mJ/cm$^2$, coinciding with a rapid increase in the emission intensity, and (iii) stimulated emission for $I_P \gtrsim 0.3$ mJ/cm$^2$. In particular, the variation of experimental emission intensity *vs.* $I_P$ is amenable to a close fit to a solution of the rate equation (Figure 3D), yielding a spontaneous emission factor $\beta = 0.03 \pm 0.01$ and laser threshold energy density $I_{th} = 0.22 \pm 0.03$ mJ/cm$^2$ (Supplementary Section VIII). The uncertainties in $\beta$ and $I_{th}$ are one standard deviation of the nonlinear fit parameters of the measured data to the laser rate equation. Moreover, the emission linewidth rapidly decreases as a function of increasing $I_P$, reaching a minimum value $\Delta = 0.28 \pm 0.15$ nm – a record-narrow linewidth for any room-temperature plasmon laser reported to date (*1, 9*), and furthermore remaining relatively constant for higher values of $I_P$. The uncertainty in the measurement of the resonance linewidth is one standard deviation of the Lorentzian fit parameter.

Further evidence to support lasing behavior of the dye-coated SPP cavity-resonator is provided by consistency with predictions of FDTD simulations exploiting a four-level model for the gain medium (Supplementary Section VII). The FDTD modelled evolution of $I_E$ vs. $I_P$, displaying a canonical 'kink' shape on a log–log scale, is consistent with three distinct emission regimes characteristic of a standard laser as it transitions through threshold (Figure 3E), and closely matches in shape the experimentally measured transition through threshold for SPP lasing (Figure 3C). The FDTD simulated magnetic field (at peak emission wavelength $\lambda_E = 630$ nm) resulting from pumping of the gain-medium-decorated cavity with an above-threshold pump-intensity $I_P = 2 \times 10^6$ W/cm$^2$ (at $\lambda_P = 480$ nm) is shown in figure 3F. The magnetic field amplitude $H_y$ is characterized by a laterally trapped SPP standing wave exhibiting twelve field-nodes (instead of the eight nodes observed in the case of the empty passive resonator, due to wave compression resulting from the background refractive index of the gain medium). Coincidence between the chosen slit position $\zeta' = 55$ nm and a node of the magnetic field (the node closest to the cavity center) is achieved as designed.

In summary, we achieve visible frequency ultra-narrow linewidth plasmon lasing at room-temperature by leveraging SPPs propagating in a Fabry-Perot cavity. Transmission-configuration sampling of the lasing SPP mode is achieved via a low-perturbation evanescently coupled recessed nanoslit. The open-cavity configuration intrinsically allows for efficient interaction between an analyte and the evanescent tail of the narrow-linewidth SPP lasing mode extending into the open space above the gain medium (Supplementary Section IX), hinting at promising applications as an integration-friendly, high-figure-of-merit platform for biological, chemical and environmental sensing.

**Acknowledgments:** W. Z., T. X. and A. A. acknowledge support under the Cooperative Research Agreement between the University of Maryland and the National Institute of Standards and Technology Center for Nanoscale Science and Technology, Award#70NANB10H193, through the University of Maryland. T. X. acknowledges support from the Thousand Talents Program for Young Professionals, Collaborative Innovations Center of Advanced Microstructures and the Fundamental Research Funds for the Central Universities.

**Figure Captions**

**Fig. 1. Design of SPP Fabry-Perot cavity.** (**A**) Schematic diagram of a plasmonic open-cavity resonator, consisting of an opaque Ag film forming two vertical Ag sidewalls (separation $l$) and a horizontal floor, and illuminated with white-light from free-space. The schematic standing wave (red) illustrates trapping of a resonant SPP mode. A recessed subwavelength-width slit, offset from cavity center by distance $\zeta$, evanescently samples the resonant SPP mode and transmits a wave of proportional intensity, $I_t$, into the far-field via a transparent substrate. (**B**) FDTD simulations of reflectance $R$ of an SPP mode impinging on an infinitely tall, perfectly flat Ag sidewall vs. angular deviation of sidewall from the surface normal, $\Delta\theta$ and free-space wavelength, $\lambda_0$. (**C**) FDTD simulations of cavity quality factor $Q$ vs. slit recess $\delta$ and end-width $\sigma$, assuming slit-sidewall taper angle = 15°, cavity length $l$ = 2.5 μm, vertical mirror height $h$ = 2 μm, free-space wavelength $\lambda_0$ = 650 nm. (**D**) Simulated $Q$ (blue circles) along with sinusoidal fit (dashed blue line) vs. $\zeta$, assuming cavity geometry of **C** along with $\delta$ = 50 nm and $\sigma$ = 10 nm. Surface magnetic field intensity $|H_y|^2$ at each slit position $\zeta$ is displayed for reference (solid orange line). The uncertainties for the simulated $Q$ are one standard deviation based on propagation of uncertainty from the Lorentzian curve-fitting of the resonance linewidths. (**E**) FDTD simulated normalized $H_y$-field distribution at $\lambda_0$ = 647 nm assuming cavity geometry of **D** for slit position $\zeta$ = 465 nm. For ease of visualization $H_y$ is multiplied by 100 in the plane below the cup-floor ($z < 0$). (**F**) Simulated spectrum of the transmitted light intensity sampled at a distance 500 nm below the exit aperture of the slit.

**Fig. 2. Fabrication and optical characterization of the SPP cavity.** (**A**) Silicon mesa used for the template-stripping method to fabricate the cup resonator with tall sidewall height and small sidewall taper angles. (**B**) Scanning electron microscope (SEM) image of focused-ion-beam (FIB) cross-section of the fabricated trough-shaped resonator, exposing the profile of the fabricated recessed slit underneath the cavity floor. **Inset:** magnified view of cross-section of recessed slit. (**C**) Measured spectrum of the transmitted light intensity out-coupled by the slit aperture, $I_t$, under illumination of the open-side of the cavity with incoherent white-light. The measured spectrum is fitted with a Lorentzian shape to estimate the linewidth $\Delta$ and quality factor $Q$.

**Fig. 3. Experimental characterization of the open-cavity SPP laser.** **(A)** Schematic diagram of an open-cavity SPP laser, consisting of a gain-medium decorated open-cavity resonator, illuminated with a pump-beam (green) of intensity $I_P$. The schematic standing wave (red) illustrates a trapped SPP lasing mode. A recessed slit, offset from cavity center by distance $\zeta'$, evanescently samples the lasing SPP mode and transmits a wave of proportional intensity, $I_E$, into the far-field. **(B)** Evolution of the spectrum of $I_E$ with increasing pump intensity $I_P$. **Inset:** SEM image of FIB cross-section of the fabricated SPP laser. A thin layer of platinum is deposited on top of the device to assist with FIB cross-section. The orange dashes outline the spin-coated gain medium. **(C)** Emission linewidth, $\Delta$, and **(D)** Log-scale light-light curve of the out-coupled light emission, $I_E$, vs. $I_P$. The uncertainties in the measurements of $I_P$ are one standard deviation based on measurement of pump energy from 100 consecutive pulses. **Inset of (C):** $I_E$ vs. $I_P$ on a linear scale indicating the lasing threshold $I_{th}$. **(E)** FDTD simulated log-scale light-light curve depicting the intensity of the slit out-coupled light emission, $I_E$ (at $\lambda_E = 630$ nm), vs. pump intensity $I_P$ (at $\lambda_P = 480$ nm). **(F)** FDTD simulated normalized $H_y$-field distribution at $\lambda_E = 630$ nm for pumping at $\lambda_P = 480$ nm with intensity $I_P = 2 \times 10^6$ W/cm$^2$. For ease of visualization $H_y$ is multiplied by 100 in the plane below the cup-floor ($z < 0$).

# Fig. 1. Design of SPP Fabry-Perot cavity

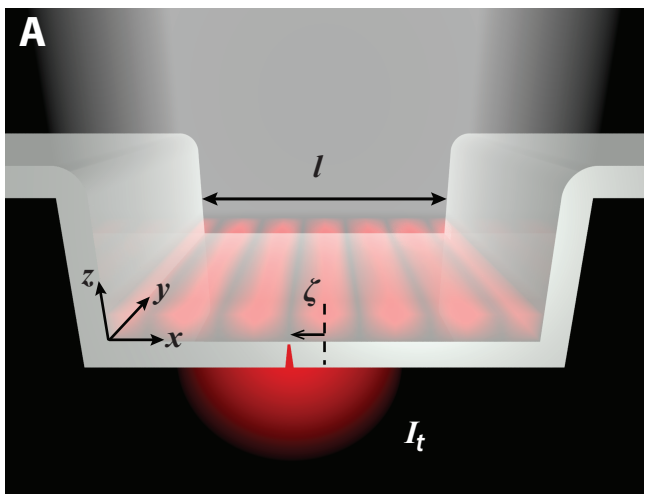

**A**

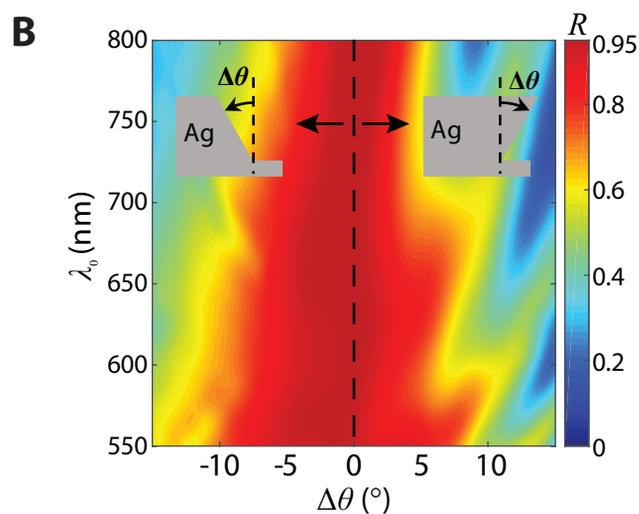

**B**

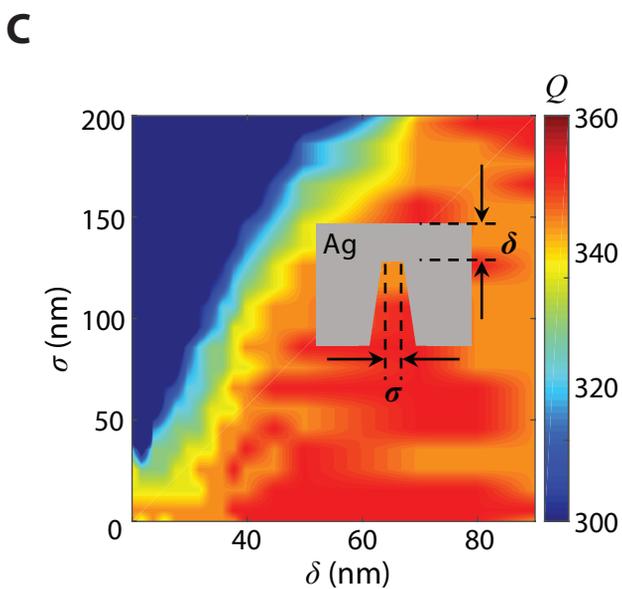

**C**

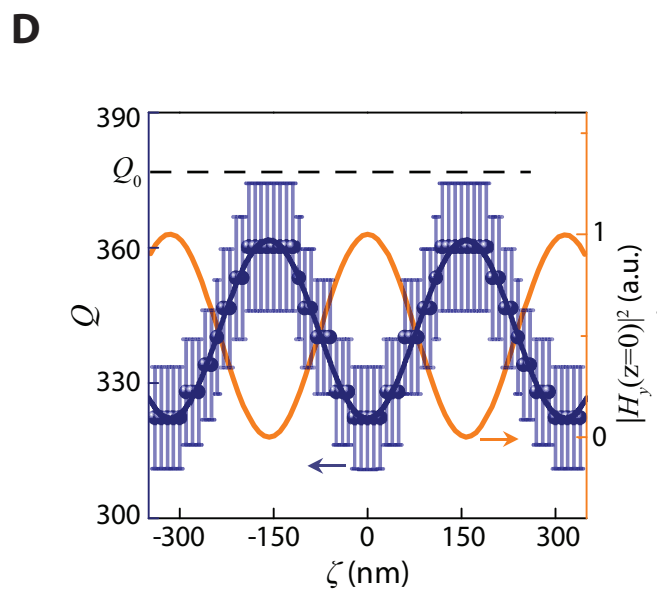

**D**

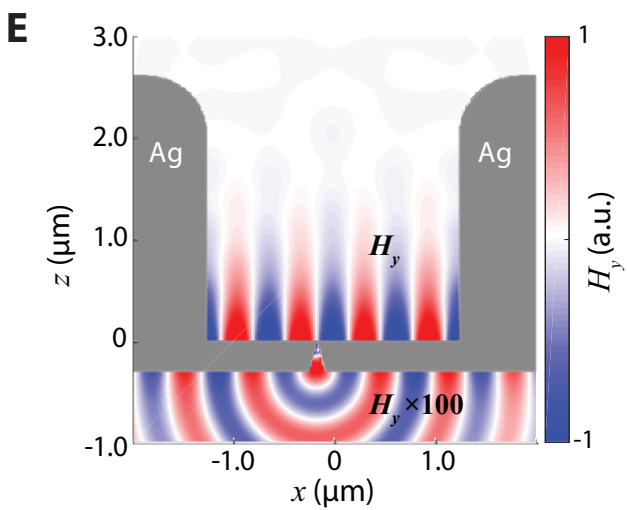

**E**

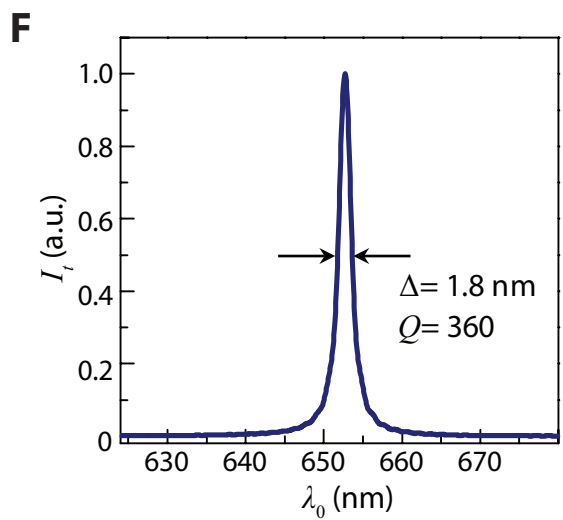

**F**

Fig. 2. Fabrication and optical characterization of the SPP cavity.

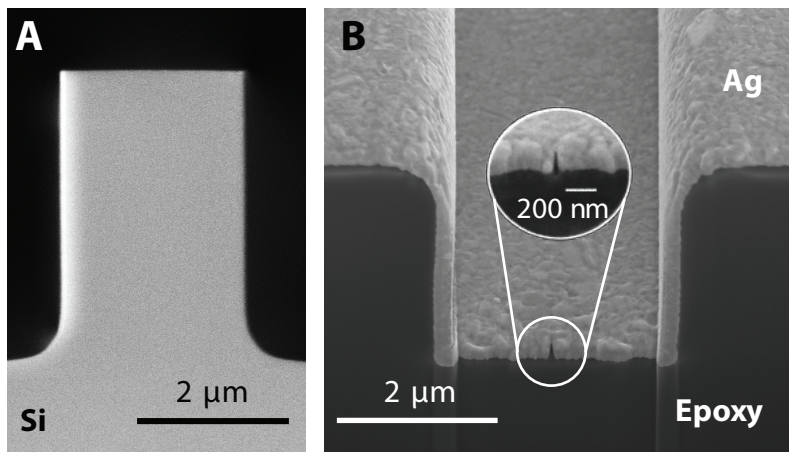

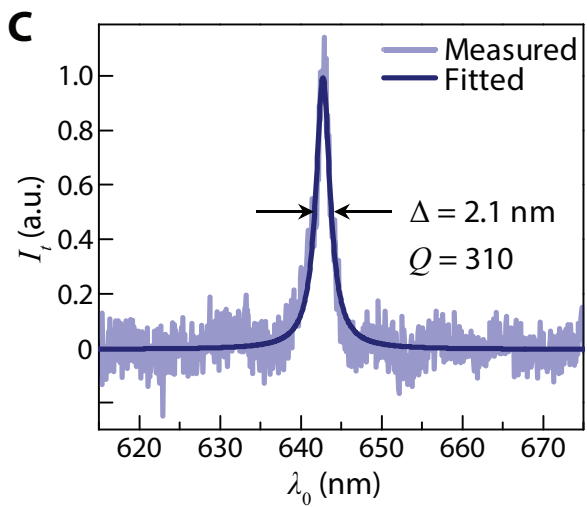

Fig. 3. Experimental characterization of the open-cavity SPP laser.

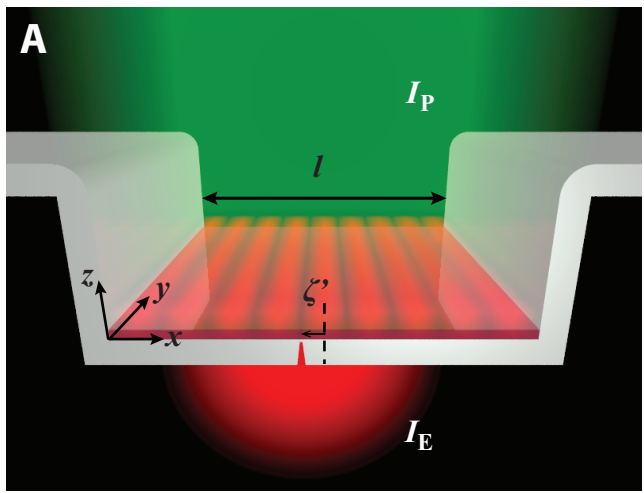
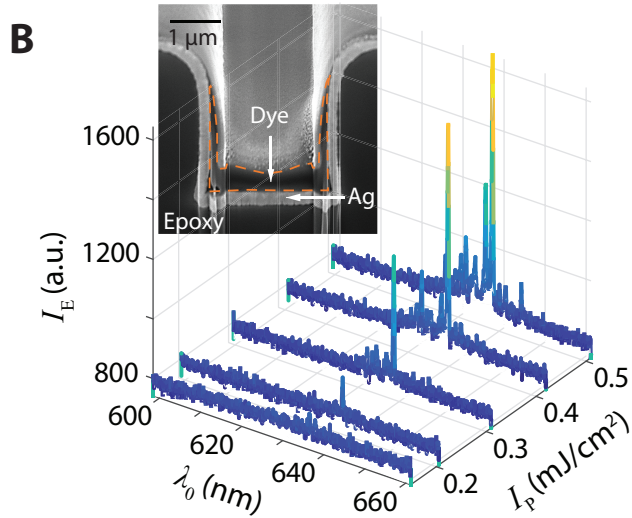
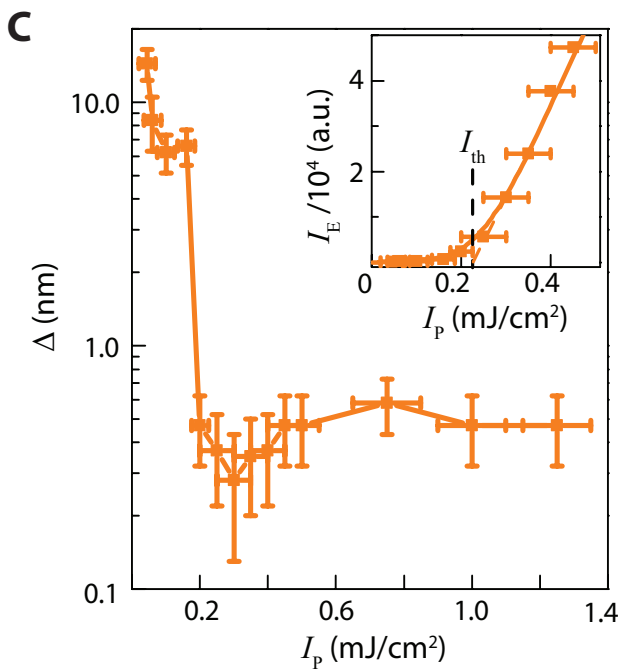
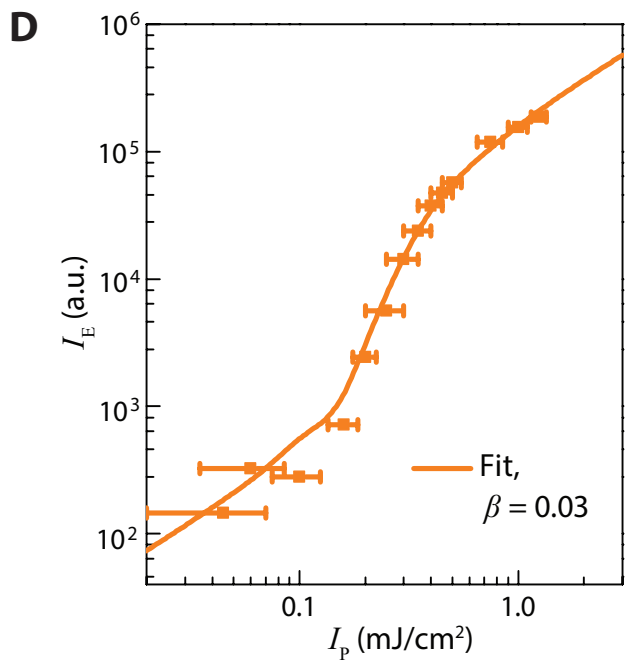
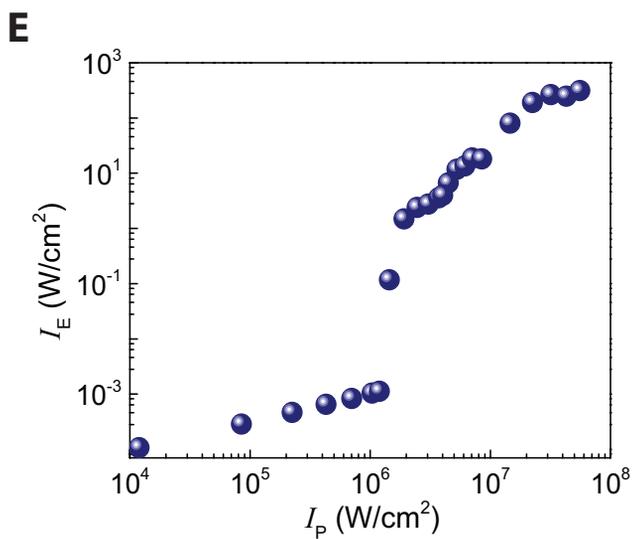
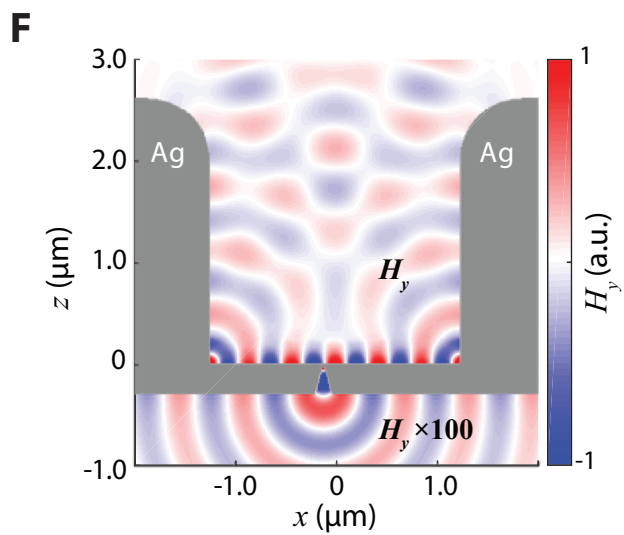

# Supplementary Information for

# Surface-Plasmon-Polariton Laser based on an Open-Cavity Fabry-Perot Resonator

Wenqi Zhu[1,2], Ting Xu[1,2,†], Amit Agrawal[1,2] and Henri J. Lezec[1,*]

[1]Center for Nanoscale Science and Technology, National Institute of Standards and Technology, Gaithersburg, MD 20899, USA.

[2]Maryland Nanocenter, University of Maryland, College Park, MD, 20742 USA.

[†]Current address: National Laboratory of Solid State Microstructures, College of Engineering and Applied Sciences and Collaborative Innovation Center of Advanced Microstructures, Nanjing University, 22 Hankou Road, Nanjing 210093, China.

*Corresponding author. E-mail: henri.lezec@nist.gov

# I. Nanofabrication of open-cavity resonator and laser structures

Fabrication of the open-cavity resonator for surface plasmon polaritons (SPPs) was performed using the template stripping method (*S1*) with a pattern template consisting of a reusable, high-aspect-ratio Si mesa having a flat ultra-smooth top surface and near-vertical sidewalls.

As a first step in the fabrication of the Si mesa template, a layer of 50 nm-thick $Si_3N_4$ was deposited onto a silicon wafer via plasma-enhanced chemical vapor deposition. A 300-nm-thick spin-coated layer of poly-methyl methacrylate (PMMA) resist was then exposed with an inverse cavity pattern using electron-beam lithography at 100 keV, developed for 60 s in methyl isobutyl ketone (MIBK), and rinsed for 30 s in isopropyl alcohol (IPA). Room-temperature reactive ion etching and cryogenic (-110 °C) deep-silicon etching (*S2*) were used consecutively to etch the $Si_3N_4$ layer and Si, forming mesas of width $l \approx 2.5$ μm and height $h \approx 2.5$ μm. The wafer was then sequentially soaked in acetone and a nitride etchant (180 °C) to remove the PMMA resist and the $Si_3N_4$, respectively. The sidewall roughness of the Si mesa was effectively reduced by growing a 300-nm-thick layer of $SiO_2$ using an oxidation furnace (*S3*), and subsequently removing the layer using hydrofluoric acid (HF). Template fabrication was then finalized by exposing the Si mesa to a "piranha" etch solution to render it weakly adhesive to evaporated layers of metals such as Ag.

As a first step in the formation of the open-cavity resonator, Ag is angle evaporated onto the rotating Si template (at 70° with respect to the sample normal), to top-surface and sidewall thicknesses of ≈ 300 nm and ≈ 250 nm, respectively. A tapered-profile slit of subwavelength width is then patterned through the top-surface Ag-film, parallel to the mesa sidewalls, using low-current (1 pA), high resolution focused-ion-beam (FIB) milling, stopping short of the Ag-Si interface with a recess distance ≈ 50nm. The resulting slit width is ≈ 50 nm in the plane of the film and $\sigma \approx 10$ nm at its deepest point (as characterized by FIB cross-section analysis of a reference slit milled under identical conditions). Following application of optical-grade transparent epoxy, backing with a glass slide and thermal curing of the epoxy, mechanical stripping of the slide from the Si template then yields the final Ag open-cavity structure, decorated with a subwavelength probing slit recessed below the cavity floor by distance $\delta \approx$ 50nm.

To transform the passive resonator into an active SPP lasing device, a thin layer of solid-state gain medium consisting of 3 mM dye (DCM) – doped PMMA is spin-coated onto the Ag floor of the open-cavity, resulting in a nominal thickness at cavity center of $\approx$ 260 nm.

## II. Optical characterization of open-cavity SPP resonator and SPP laser

For passive cavity characterization, the samples were illuminated from the open-side of the cavity with a halogen lamp (wavelength range: 450 nm to 850 nm). Light transmitted through the sampling slit was collected using 100× microscope objective ($NA = 0.75$) and directed to a triple-grating spectrometer connected to a cooled Si-CCD camera. Typical camera integration time was 30 s. A 150 grooves/mm grating was used to measure the broadband spectra and to find the resonance wavelength. A 600 grooves/mm grating that provides high spectral resolution was then used to characterize the corresponding resonance peak, from which the full-width half-maximum (FWHM) linewidth and the quality factor $Q$ are determined. For plasmon laser characterization, a procedure similar to that used for passive characterization is followed, except that the optical source is replaced by a tunable nanosecond pulsed laser ($\lambda_P$ = 480 nm, repetition rate = 10 Hz, pulse width $\approx$ 5 ns). A 514 nm long-pass filter is placed in front of the spectrometer entrance to block the residual pump light. Typical camera integration time for these measurements was 10 s.

## III. Morphological and optical characterization of template-stripped Ag surface

Template-stripped Ag surfaces have been shown to exhibit absorption-induced SPP propagation losses that are smaller than those of as-deposited Ag surfaces (*S1*). The evaporated template-stripped Ag surfaces used in this study exhibit typical root-mean-square roughness of ≈ 2.9 nm as measured by an atomic force microscope (Figure S1). The corresponding SPP propagation decay length $L_x$ is experimentally measured (using the method described in ref. *S1*) to be ≈ 25 μm at $\lambda_0$ = 650 nm (Figure S2, blue spheres), a value that closely matches the theoretical SPP decay length calculated using the bulk effective permittivity of template-stripped Ag measured by a spectroscopic ellipsometer (Figure S2, dashed black line).

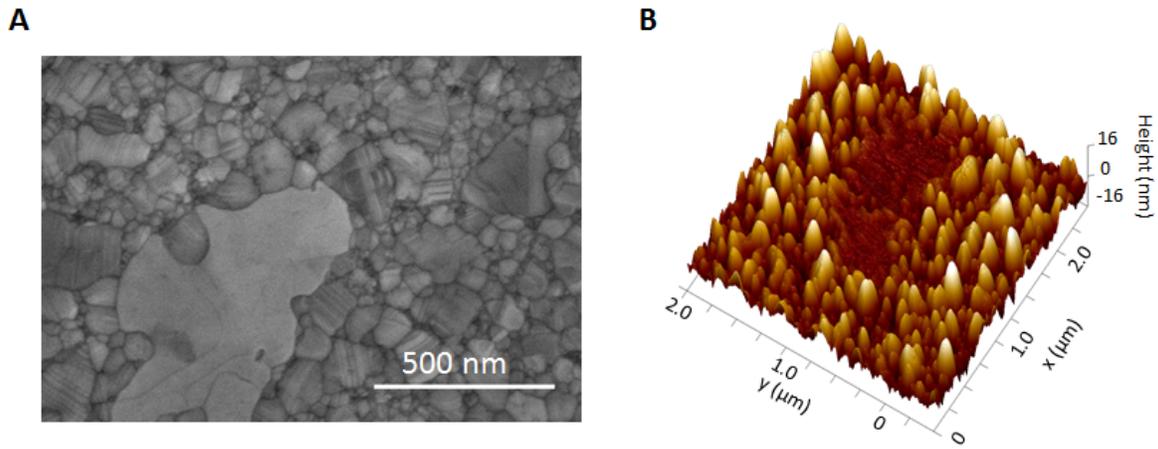

**Figure S1. Surface morphology characterization of template-stripped Ag film. (A)** Top-down scanning electron microscope (SEM) image and **(B)** atomic force microscope (AFM) image of a template-stripped Ag-air interface. AFM measurements yield a root-mean-squared roughness of ≈ 2.9 nm and an arithmetic-average roughness of ≈ 2.1 nm.

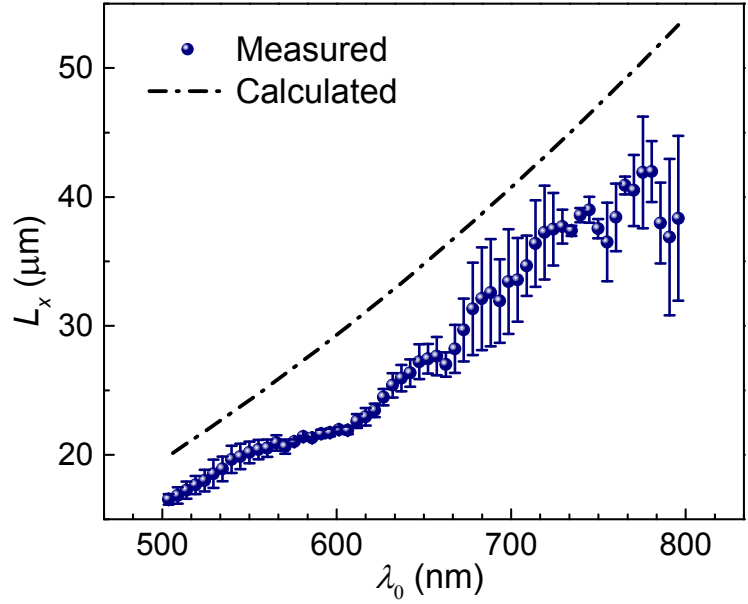

**Figure S2. Propagation decay length of SPPs propagating on a template-stripped Ag-air interface.** Experimentally measured $1/e$ decay length $L_x$ of SPPs for free-space wavelengths ranging from 500 nm to 800 nm (blue spheres). The theoretical SPP decay length calculated using the bulk effective permittivity of template-stripped Ag (dashed black line).

## IV. Numerical simulation of cavity-sidewall reflectance for SPPs

FDTD simulations at $\lambda_0 = 650$ nm, assuming perfectly vertical sidewalls ($\Delta\theta = 0°$), are used to study the dependence of sidewall SPP reflectivity $R$ on sidewall height $h$. Simulations show that $R$ reaches values within 1 % of its asymptotic limit $R_\infty \approx 0.95$, for sidewall heights $h > 1.5$ μm (Figure S3), in other words for heights exceeding by a factor of $\approx 4$ the corresponding SPP decay length given by simulation ($d \approx 400$ nm).

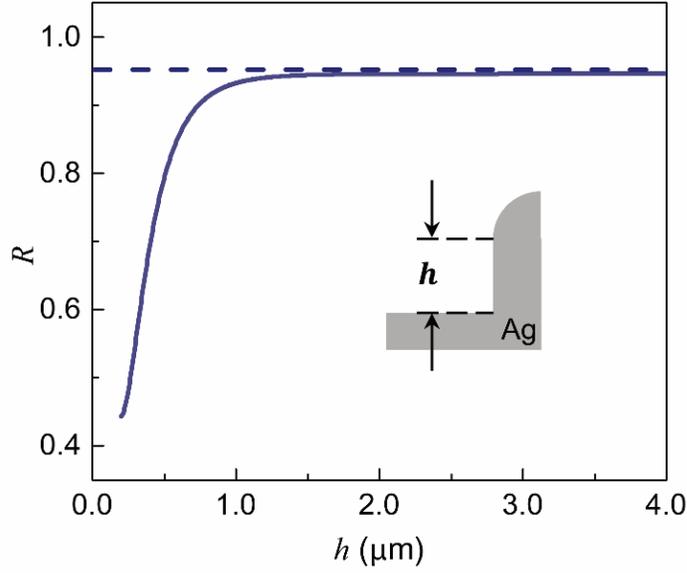

**Figure S3. Sidewall SPP reflectance $R$ as a function of sidewall height $h$.** The dashed blue line shows the SPP reflectance for an infinitely tall, perfectly flat and vertical Ag sidewall, $R_\infty \approx 0.95$. $R$ asymptotically increases towards $R_\infty$ as a function of increasing $h$, reaching values within 1 % of $R_\infty$ for $h > 1.5$ μm.

**V. Quality factor and Finesse of a lossy Fabry-Perot cavity for SPP modes**

The quality factor $Q$ and Finesse $\mathcal{F}$ of a slit-free SPP Fabry-Perot cavity of length $l$, with an anticipated resonance frequency $\nu_0$ (or free-space wavelength $\lambda_0$), can be estimated by using the lossy Fabry-Perot cavity model (*S4*). Consistent with a free-space Fabry-Perot resonator, the resonance condition for a resonator utilizing SPPs propagating along a metal cavity floor enclosed by a pair of reflective sidewalls is reached when the cavity can trap an integer numbers of SPP half wavelengths, i.e. for cavity lengths:

$$l = m \frac{\lambda_{SPP}}{2},$$

where $m$ is a positive integer (1, 2, ...), and $\lambda_{SPP}$ is the SPP wavelength (Figure S4A). At a free-space wavelength $\lambda_0$, the wavelength of an SPP mode propagating along a Ag-air interface is given by $\lambda_{SPP} \approx \lambda_0/n_{SPP}$, where

$$n_{SPP} = \text{Re}\left[\sqrt{\frac{\varepsilon_{Ag}}{\varepsilon_{Ag}+1}}\right], \quad (S1)$$

is the SPP mode index, and $\varepsilon_{Ag}$ is the complex-dielectric permittivity of the template-stripped Ag-film at $\lambda_0$.

The total SPP round-trip loss in an open cavity resonator results from a combination of: (a) the absorption and scattering induced propagation loss of SPPs along the cavity floor, and (b) the absorption and scattering induced SPP reflection loss at each of the sidewalls. The propagation and reflection losses together limit the highest quality factor $Q$ and Finesse $\mathcal{F}$ achievable using a Fabry-Perot cavity. For one round-trip propagation within a cavity of length $l$, the SPP power exponentially attenuates as $\exp(-2\alpha_s l)$, where $\alpha_s = 1/L_x$ is the SPP propagation loss-coefficient (where $L_x$ is plotted in figure S2). Combining propagation loss with the SPP reflectance $R$ at each sidewall (Figure 1B), the overall round-trip SPP attenuation is:

$$r^2 = R^2\exp(-2\alpha_s l) = \exp(-2\alpha_r l),$$

where $\alpha_r$ is the effective overall distributed-loss coefficient given by:

$$\alpha_r = \alpha_s + \frac{1}{2l}\ln\frac{1}{\mathcal{R}^2}.$$

The finesse $\mathcal{F}$ of the SPP Fabry-Perot resonator (*S4*) is expressed as:

$$\mathcal{F} = \frac{\pi\exp(-\alpha_r l/2)}{1 - \exp(-\alpha_r l)},$$

and the quality factor $Q$ is expressed as:

$$Q = \frac{\nu_0}{\delta\nu} = \frac{c/\lambda_0}{c\alpha_r/2\pi},$$

where $\delta\nu = c\alpha_r/2\pi$ is the FWHM linewidth of the resonance and $c$ is the speed of light in vacuum. The Fabry-Perot resonators used in this study are fabricated using Ag for the metal surfaces of the cavity and targeted to operate at a nominal resonant wavelength $\lambda_0 = 650$ nm, corresponding to a SPP wavelength $\lambda_{SPP} = 625$ nm (calculated using equation S1). The quality factor $Q$ and Finesse $\mathcal{F}$ values for possible cavity-lengths $l$, corresponding to an integer multiple of $\lambda_{SPP}/2$, that support a resonance at $\lambda_0 = 650$ nm are shown in figures S4B and S4C, respectively. These results clearly depict the tradeoff between quality factor $Q$ and Finesse $\mathcal{F}$ for increasing cavity length $l$, where $Q$ ($\mathcal{F}$) increases (decreases) with increasing $l$. For simplicity, we define a cavity figure-of-merit $\Gamma = Q * \mathcal{F}$ (Figure S4D) that exhibits a maximum value ($\Gamma = 6750$), corresponding to a simultaneously high value of quality factor and Finesse, respectively $Q = 375$ and $\mathcal{F} = 24$, for a cavity length $l = 2.5$ μm ($m = 8$).

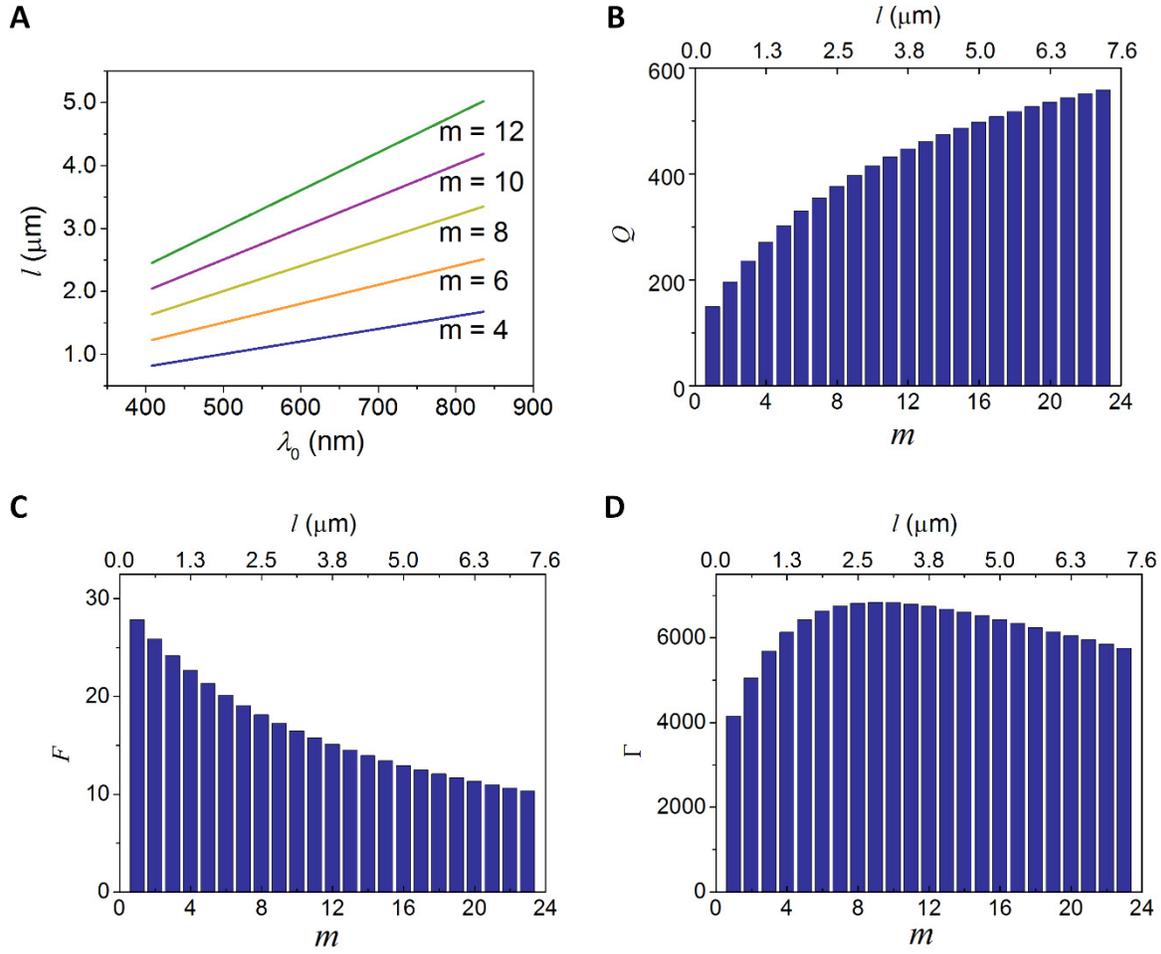

**Figure S4. Design of SPP resonator based on a lossy Fabry-Perot-cavity model. (A)** Cavity length corresponding to the trapping of an integer numbers (*m*) of SPP half wavelengths for free-space wavelength ranging from 400 nm to 850 nm. Variations in **(B)** Quality factor $Q$, **(C)** Finesse $\mathcal{F}$ and **(D)** Figure-of-merit $\Gamma$ as a function of increasing cavity length $l$ (or mode number $m$) for a cavity supporting a resonance at $\lambda_0$ = 650 nm vs. $l$ (or $m$) for a pre-determined resonance wavelength. Simultaneously high values of $Q$, $\mathcal{F}$ and $\Gamma$ are achieved for cavity length $l$ = 2.5 µm ($m$ = 8).

**VI. FDTD simulation-based optimization of open-cavity SPP resonator**

The geometrical parameters of open-cavity SPP resonators are optimized for operation at and around $\lambda_0$ = 650 nm using finite-difference-time-domain (FDTD) numerical simulations. The complex permittivity values employed in the simulations to model the cavity floor and sidewalls were obtained from spectroscopic ellipsometry measurements of template-stripped Ag. A resonance at $\lambda_0$ = 647 nm with a quality-factor $Q_0$ = 360 is achieved via simultaneous optimization of the cavity length ($l$ = 2.5 μm), sidewall height ($h$ = 2 μm), sidewall angle ($\theta$ = 90°), slit recess ($\delta$ = 50 nm), minimum slit-width ($\sigma$ = 10 nm), slit sidewall taper angle (15°) and slit offset from cavity center $\zeta$ = 155 nm (Figures 1 B-D).

**VII. FDTD simulation of open-cavity SPP laser**

FDTD simulations were used to demonstrate SPP lasing from an open-cavity SPP resonator incorporating a layer of gain medium on the cavity floor and optically pumped from the cavity side. In the FDTD model, the cavity length, sidewall heights and slit-profile are set to the same values as those chosen for the optimized passive SPP resonator (Supplementary Section VI). The slit offset from cavity center, $\zeta'$, is however modified to coincide with a simulated node of the magnetic field of the SPP standing wave (see below). The simulated dye-layer thickness of 260 nm is chosen to match the dye-layer thickness at cavity center obtained upon spin coating of the experimental device (Figure 3B). The gain medium is treated as a four-level system (*S5*) with emission and absorption cross-sections exhibiting linewidth of ≈ 10 nm and centered at $\lambda_E$ = 630 nm and $\lambda_P$ = 480 nm, imbedded to a concentration N = 2 × 10$^{18}$ cm$^{-3}$ in host medium of relative dielectric constant 2.25 (mimicking PMMA). The cavity is pumped from its open-side at normal incidence using a continuous-wave source at $\lambda_P$ = 480 nm. The intensity of the transmitted light ($I_E$), out-coupled by the slit, and radiated on the substrate-side of the cavity is recorded at a distance of 500 nm below the slit as a function of increasing pump-power ($I_P$). The FDTD modelled evolution of $I_E$ vs. $I_P$, displaying a canonical 'kink' shape on a log–log scale, is consistent with three distinct emission regimes characteristic of a standard laser as it transitions through threshold (Figure 3E), and closely matches the experimentally measured SPP lasing behavior (Figure 3D). The FDTD simulated magnetic field $H_y$ of the lasing cavity above-threshold at the emission wavelength ($\lambda_E$ = 630 nm, Figure 3A) is characterized by a laterally trapped SPP standing wave exhibiting twelve field-nodes (instead of the eight nodes observed in the case of the empty passive resonator), due to wave compression by a factor ≈ 1/1.5 resulting from the background refractive index of the gain medium ($n \approx 1.5$). The recessed slit location is correspondingly chosen to coincide with the first magnetic-field node of the SPP standing wave at the emission wavelength, located at an offset $\zeta$ = 55 nm from the cavity center.

## VIII. Fitting of lasing rate equation to experimental SPP emission

A simplified rate-equation model is adopted to study the dynamics of the photon–exciton interaction in the SPP laser cavity (*S6*). The time-evolution of the exciton density *n* and emitted photon density *s* inside the cavity can be expressed using the set of coupled differential equations:

$$\frac{dn}{dt} = \sigma p - An - \Gamma Asn,$$

and

$$\frac{ds}{dt} = \beta An + \Gamma Asn - \gamma s,$$

where $\sigma$ is the exciton generation efficiency, $p$ is the pump photon density, $A^{-1}$ is the exciton lifetime inside the cavity, $\Gamma$ is the gain-mode overlap factor, $\beta$ is the spontaneous emission factor, and $\gamma$ is the non-radiative photon lifetime in the cavity. The coupled rate equations are solved under the assumption that the SPP laser is pumped by a steady-state continuous-wave light, corresponding to a time-invariant pump photon density $p$. The laser emission intensity, $I_E \propto s$, at steady-state is calculated for increasing pump power, $I_P \propto p$, and the pump-emission curve for various values of $\beta$ are plotted on log-log and linear-linear scales (Figure S5A and S5B, respectively). The experimental estimate of $\beta$ in figure 3D is achieved by fitting the solution of the rate equation to the experimentally measured pump-emission light-light curve.

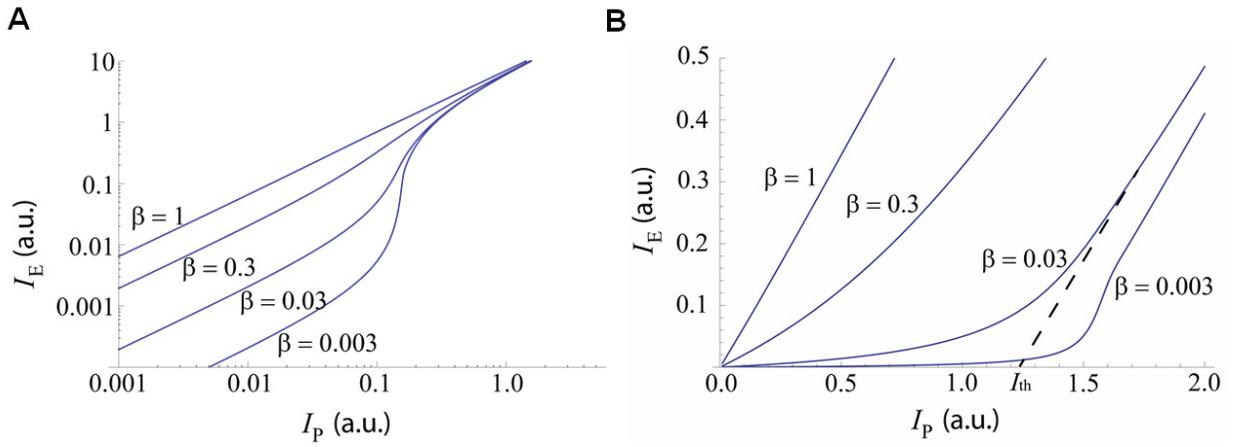

**Figure S5. Laser emission as a function of varying spontaneous emission factor.** The emission intensity ($I_E \propto s$) as a function of the pump power ($I_P \propto p$), obtained by solving the laser rate equations for various values of spontaneous emission factor $\beta$, is plotted in **(A)** log-log scale showing the canonical 'kink' behavior, and **(B)** linear-linear scale, showing the canonical lasing threshold behavior.

### IX-A. Refractive index sensing using the passive SPP resonator

To evaluate the sensitivity of the plasmonic cavity resonator to a superficial perturbation in index of refraction, ultra-thin $Al_2O_3$ layers of index n = 1.77 and thickness ranging from 0.8 nm to 8 nm were conformally deposited onto the surfaces of the cavity using atomic layer deposition.

Nanometer-scale spectral shifts of resonance peak $\lambda_P$ to longer wavelengths, as a function of increasing layer thickness, are easily resolvable (Figure S6) due to the narrow resonance linewidth characteristic of the passive SPP resonator. The experiments yield a refractive index sensitivity $S = d\lambda_p/dn \approx 714$ nm·RIU$^{-1}$ along with an FOM = $S / \Delta = dQ/dn \approx 340$. Though comparable to that of a conventional SPR sensor in a Kretschmann configuration, this FOM is achieved with a lateral device footprint which is significantly smaller.

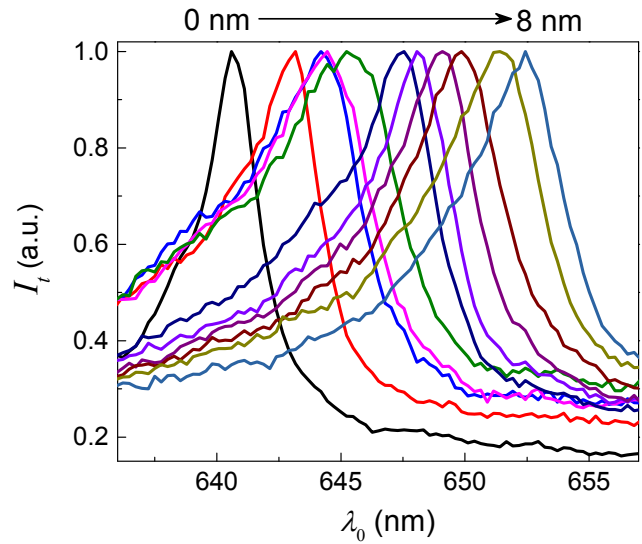

**Figure S6. Refractive index sensing using passive SPP resonator.** Experimentally measured spectra of the out-coupled light from the SPP resonator as a function of increasing Al$_2$O$_3$ layer thickness, varying from 0.8 nm to 8 nm.

**IX-B. Refractive index sensing using the active SPP lasing mode**

The performance of a refractive index sensor based on interaction of an analyte with the evanescent tail of an active SPP lasing mode extending into the open space above the gain medium (Figure S7A) is explored using FDTD simulations. This is accomplished by changing the refractive index of the medium immediately above the gain media within the laser cavity from n = 1.00 to n = 1.04, and observing the shift in the wavelength of the lasing mode (Figure S7B). The simulations yield a refractive index sensitivity S = $d\lambda_p/dn$ = 250 nm·RIU$^{-1}$, a value that is approx. a factor of three smaller than that of the passive resonator. However, since the linewidth of the lasing mode (Δ ≈ 0.2 nm) is approx. a factor of ten smaller than that of the passive resonator, the refractive index sensor based on the lasing mode yields a record theoretical FOM ≈ 1250, suggesting applications as an open-cavity SPP laser based high-FOM refractive index sensor.

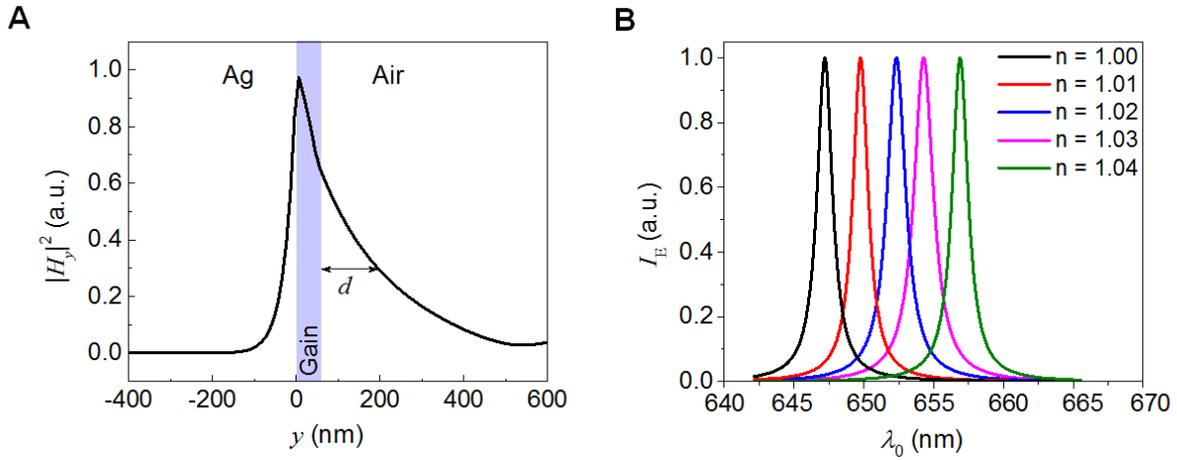

**Figure S7. Refractive index sensing using the active SPP lasing mode. (A)** Surface magnetic field intensity, $|H_y|^2$, at the center of the cavity (ζ = 0 nm) depicting the exponential decay of the lasing SPP mode into the Ag film ($y < 0$), and the double-exponential decay above the Ag film ($y > 0$). The thickness of gain-media used in the simulations was 50 nm. **(B)** Spectral shift of the SPP lasing mode as a function of change in the refractive index of the dielectric medium above the gain media of the cavity.